# Slipping friction of an optically and magnetically manipulated microsphere rolling at a glass-water interface

Rodney Ray Agayan

Department of Chemistry, University of Michigan, Ann Arbor, Michigan 48109-1055

and

Applied Physics Program, University of Michigan, Ann Arbor, Michigan 48109-1120

Ron Gordon Smith

Department of Chemistry, University of Michigan, Ann Arbor, Michigan 48109-1055

Raoul Kopelman

Department of Chemistry, University of Michigan, Ann Arbor, Michigan 48109-1055

and

Applied Physics Program, University of Michigan, Ann Arbor, Michigan 48109-1120

**Abstract**

The motion of submerged magnetic microspheres rolling at a glass-water interface has been studied using magnetic rotation and optical tweezers combined with bright-field microscopy particle tracking techniques. Individual microspheres of varying surface roughness were magnetically rotated both in and out of an optical trap to induce rolling, along either plain glass cover slides or glass cover slides functionalized with polyethylene glycol. It has been observed that the manipulated microspheres exhibited nonlinear dynamic rolling-while-slipping motion characterized by two motional regimes: At low rotational frequencies, the speed of microspheres free-rolling along the surface increased proportionately with magnetic rotation rate; however, a further increase in the rotation frequency beyond a certain threshold revealed a sharp transition to a motion in which the microspheres slipped with respect to the external magnetic field resulting in decreased rolling speeds. The effects of surface-microsphere interactions on the position of this threshold frequency are posed and investigated. Similar experiments with microspheres rolling while slipping in an optical trap showed congruent results.

**Introduction**

The study of micro- and nanotribology has been gaining great interest. In particular, the knowledge of effects such as friction, adhesion, and lubrication at submerged surfaces, on the micro- and nanoscale, is crucial for the development of many cell sorting and separation devices that are based on micro-electro-mechanical systems (MEMS) and microfluidic chips [1, 2]. In addition, such knowledge can help us to better understand and potentially mimic the mechanisms of locomotion of some cellular biological systems [3].

Surfaces can greatly affect the hydrodynamic motion of nano- and microscale objects. Optically torqued nanorods have been shown to undergo transitions from motor to rocking behavior due to interactions with a nearby surface [4]. Wax microdisks that are optically trapped at an opposing wall can exhibit switchback oscillations representative of a Hopf bifurcation [5]. The physical properties of the surfaces themselves can also be engineered to provide specific adhesive and frictional qualities through nanopatterning [6, 7]. In numerous biological systems of interest, the dynamics of locomotion depend on surface-cell interactions. For example, the rolling velocities of white blood cells are mediated by density and binding affinity, among numerous other properties of selectin-coated surfaces [8, 9]. The oscillatory angular motion of magnetotactic bacteria in rotating magnetic fields can also be affected by the rotational drag near a surface [10, 11].

In this study, we report on a simple method for comparing the frictional properties of submerged surfaces on the micron scale. By observing changes in the motion of a magnetic microsphere that is slipping while rotating on a planar surface, we have differentiated between untreated glass substrates and substrates with polymeric coatings. In addition, the effects of particle surface roughness were investigated.

The motion of a magnetic microparticle's rotational slipping near a glass surface was studied both for particles that are freely rolling along the surface as well as particles spatially confined by optical tweezers. Without the optical trap, particles rolled, while slipping, along the surface at different velocities, depending on the rotation frequency of an external magnetic field. For low frequencies, the velocity of locomotion-while-slipping increased with the rotation rate. Beyond a certain frequency threshold, the rolling velocity of the microsphere would decrease. This sharp change in motional behavior was often accompanied by a slight change in the overall direction of rolling. With the laser turned on, the particles were trapped but, while slipping, they rolled towards one side of the potential well created by the light intensity. Increasing the magnetic rotation rate further tended to shift the particle even farther from the trap center until a threshold was reached after which the particle did not escape, usually, but rather remained confined, while continually slipping against the glass surface, and residing closer to the trap center. Similar but symmetrically opposite motion occurred using magnetic rotation in the opposite direction. Changes in the angular direction of motion similar to those for free-rolling microspheres were also observed for trapped microspheres at high rotation rates.

Drastic changes in the rotational motion of micro-objects are characteristic of the behavior of overdamped driven nonlinear oscillators [4, 12-15]. Such systems exhibit two classes of motional behavior: (1) linear phase-locked rotation at low external driving frequencies and (2) nonlinear phase-slipping rotation at high external driving frequencies. In the rolling-while-slipping systems described here, such nonlinear behavior contributes at least partially to the motion we've observed. Additional effects due to surface-microsphere interactions are proposed and investigated.

## Experimental Details

### Sample Preparation

To explore the frequency dependence of the drag coefficient near a surface, single polystyrene microspheres coated with magnetic particles (Spherotech, Inc., Libertyville,

IL), of nominally 9 ± 1 $\mu m$ diameter, were studied using bright field reflection microscopy. Two types of microspheres were used: carboxylated and amine-functionalized. The difference in surface roughness can be seen in the scanning electron micrograph (SEM) images in Figure 1. A 1:100 dilution of particle stock (1% w/v) solution in de-ionized water was prepared. To aid in preventing particles from sticking to the glass surface, as well as to each other, this solution was further mixed with 10% aqueous sodium dodecyl sulfate (SDS) in a 1:10 10% SDS:particle ratio. The final particle mixture was then magnetized for several minutes by a 1400 *Oe* field to saturation. After vortexing the solution at 3000 *rpm* for 15 *sec* to separate aggregates, about 40 $\mu L$ of the solution was inserted between two glass cover slides separated approximately by 100 $\mu m$, with double-stick tape, and sealed with vacuum grease to prevent convection. Because the ratio between the sample chamber thickness and the microsphere radius was sufficiently large [16], hydrodynamic effects due to the top glass surface were neglected.

Glass cover slides (Erie Scientific Co, Portsmouth, NH) of thickness #0 were used for the sample cell, either untreated (directly from the package) or coated with polyethylene glycol (PEG) to inhibit adhesion to the glass caused by non-specific binding of the magnetic particles. The following PEGylation procedure was adapted from the literature [17]. Cover slides were first cleaned with 10:90, 50:50, and then 90:10 mixtures of methanol and methylene chloride, consecutively, for 15 minutes each in an ultrasonic bath. The cover slides were then thoroughly rinsed with Nanopure water. After rinsing, the cover slides were immersed in a 30:70 mixture of 30% $H_2O_2/H_2SO_4$ for 30 *min*. This was followed by another rinsing and drying step, after which the cover slides were silanized by immersion in a solution of 1% 3-glycidoxypropyl trimethoxysilane in dried toluene for 24 *hours*. This was followed by an acetone wash. In order to hydrolyze the epoxide, the cover slides were placed in a 100 *mM* NaCl solution, at a pH ~4, for 3 *hours*, and then washed again with Nanopure water. Oxidative cleavage of the "diols" was accomplished by soaking the samples in a 5 *mM* $NaIO_4$ solution for 8 *hours*, followed again by rinsing and drying with Nanopure water. PEGylation was accomplished by placing the cover slides in 17 *mM* PEG-amine (2000 MW) in $CHCl_3$ for 15 *min*. The excess solution was removed and the cover slides were placed in an oven at $74^o$ *C* for 40 *hours*. The PEGylated cover slides were then vigorously washed with Nanopure water and dried with nitrogen.

Verification of the presence of the PEG coating was accomplished by measuring the contact angle of a droplet of Nanopure water placed on a cover slide of each batch, using the static sessile drop technique. The presence of the PEG layer reduced wetting of the glass surface, thus exhibiting a contact angle of 59 ± 3°, compared to untreated cover slides which displayed a contact angle of 31 ± 4° (see Figure 2). This increase in contact angle for PEGylated glass agrees with measurements conducted by others in the literature [9].

**Optical and Magnetic Manipulation**

Particles were observed on an inverted microscope (Olympus IX-71) utilized for both bright field imaging and optical tweezing. To induce rolling, particles were simultaneously rotated about a single axis (*x*-axis) using an external rotating magnet located above the sample plane. Rotation rates ranged from 0-5.5 *Hz*. The amplitude of the magnetic field strength at the sample plane ranged from 5-10 *Oe*, depending on the sample being studied. Light from a xenon arc lamp in the visible range of the electromagnetic spectrum was coupled into a 100X (NA = 1.3) oil immersion (oil refractive index *n* = 1.513) objective and used to illuminate the sample. The visible light reflecting off of particles in the field of view was recollected by the objective and delivered to a CCD camera (Roper Scientific, Photometrics CoolSNAP ES, 1392x1040, ~30 frames/s) for image analysis. Because the particle systems contained magnetic material which was much more reflective than the surrounding medium, signals were significantly brighter than background scattering from the aqueous solution itself. To observe particles rotationally slipping in place, 780 *nm* laser light from a Ti:Sapphire laser in continuous wave mode was focused by the same objective using a dichroic beam splitter to create optical tweezers [18-21] capable of trapping the particle near the bottom glass surface of the sample cell. Just in front of the CCD camera, the laser light was optically filtered to prevent saturation over bright field signals reflecting from the sample.

Laser illumination intensity entering the objective ranged from 1-5 $mW/cm^2$, achieved by changing the laser power with a variable neutral density filter. Higher intensities induced stronger scattering forces that would lift particles upward off the glass surface in the *z*-direction. The optimum intensity was just below this threshold. At this optimal level, the radial trapping forces were strong enough to keep the particle confined in a localized region of the *xy*-plane. At the same time, the axial scattering radiation force [20, 21], which points in the direction of beam propagation and destabilizes three-dimensional trapping, did not overcome the force due to gravitational sedimentation, thus maintaining the particle close to the glass surface. In addition, due to aberrations arising from oil-glass and glass-water index of refraction mismatches, optical tweezing gradient forces were typically most efficient only about 8-10 $\mu m$ above the bottom glass cover slide (based on previous measurements with uncoated polystyrene microspheres).

The $z$-position of the glass surface was determined by locating the focus at which the smallest diameter reflected laser spot was observed with the CCD camera. The height of the center of each particle above the glass surface was then estimated using the focus knob of the microscope calibrated to be approximately 1 $\mu m$ between markings. Particle centers were typically 5-10 $\mu m$ above the glass, thus the particle surface was usually within a distance the size of a particle radius from the glass surface. With the laser trap turned off, systems in which the surfaces of the microsphere and glass were separated by a distance greater than 5 $\mu m$ did not roll while being rotated, and these data were thus discarded.

**Image Analysis**

Bright field images of particles collected with the CCD camera were analyzed using the Metamorph Imaging Analysis (Molecular Devices, Sunnyvale, CA) software package. From these images, the actual particle diameter was measured after the inter-pixel distance was calibrated using a USAF resolution target plate ($0.090 \pm 0.005$ $\mu m/pixel$). In addition, particle location in the $xy$-plane was tracked as a function of time.

The image in Figure 3(a) is a snapshot during one measurement in which the focus was below the equator of the particle. Small magnetic colloids can be seen on a circular region of the particle surface. Colloids outside of this ring are out of focus and thus cannot be seen. In Figure 3(b), the microscope was set to focus on the equator of the particle, thus the microsphere's edge can be seen reflecting and/or scattering the xenon lamp illumination, forming a bright circular ring. Faint concentric rings of larger diameter indicate light interference near the particle edge. This focus was approximately 8 $\mu m$ from the glass surface; thus the particle, measured to be $9.0 \pm 0.2$ $\mu m$ in diameter, was within one radius from the glass surface, but not necessarily in direct contact. The location of the laser trap is indicated by the blue circle, with a diameter corresponding to twice the size of the actual beamwaist at the focus. The magnetic field was rotated such that the particle rolled towards one end of the potential well created by the trap. The particle continued to rotationally slip, but translationally it fluctuated about a central position, as determined by tracking the $x$- and $y$-coordinates of the center of the ring.

In order to observe if the motion is symmetric with respect to the rotation direction, the results of two measurements – one for each rotation – were overlaid onto one image. The image in Figure 3(c) depicts such an overlay. All snapshots were averaged for each rotation direction, so as to obtain two blurred rings symmetrically displaced from the trap location. The axis joining the centers of the two rings was not perfectly aligned to the coordinate system of the CCD camera. This deviation shown in Figure 3(c), of about 10°, was due to a slight misalignment of the external magnet and was accounted for in the coordinate-tracking algorithm.

To quantify the rolling and slipping behavior of the microspheres, digital movies were recorded of the particle motion, while magnetically rotated either with or without the optical trap present. Particle location in the *xy*-plane was measured using the "threshold image" centroid-tracking method in Metamorph. This tracking scheme isolates bright images by applying an intensity threshold to each frame before calculating the center of mass of the bright object. The threshold was manually set such that the recognized object region did not extend far beyond the bright ring, thus minimizing background, and the central region within each ring was partially recognized as part of the object. The search region was approximately 3-4 pixels larger than the object region in both height and width. The percentage match between frames was set to between 50-75%. This percentage match can fall below this threshold if the brightness of the object changes, such as when the particle rolls across an unevenly illuminated field of view. In such cases, the threshold was readjusted to match the prior criteria. This position tracking scheme is known to provide an accuracy of about 10 *nm*, depending on the noise level [22]. After tracking, coordinates were adjusted to account for the angular deviation previously discussed.

## Theoretical Considerations

It is well known that in the low-Reynolds number regime the drag force $F$ and torque $N$ on a sphere of radius $R$ infinitely far from a surface rotating with angular velocity $\dot{\theta}$ and translating with velocity $v$ in a fluid of dynamic viscosity $\eta$ are given by

$$F_{sphere} = -\gamma_{trans} v \qquad (1)$$

and

$$N_{sphere} = -\gamma_{rot} \dot{\theta}, \qquad (2)$$

where $\gamma_{trans} = 6\pi\eta R$ and $\gamma_{rot} = 8\pi\eta R^3$ are the translational and rotational friction coefficients of a sphere, respectively. The dot-notation corresponds to differentiation with respect to $t$. As one approaches a surface, and assuming a no-slip boundary condition of the fluid at the surface, the friction coefficients increase. For small sphere-center-to-plane distances $h$, the friction coefficients should be multiplied by the corrections factors $c_{trans}$ and $c_{rot}$, respectively [23]:

$$c_{trans} = \left[1 - \frac{9}{16}\left(\frac{R}{h}\right) + \frac{1}{8}\left(\frac{R}{h}\right)^3 - \frac{45}{256}\left(\frac{R}{h}\right)^4 - \frac{1}{16}\left(\frac{R}{h}\right)^5\right]^{-1} \qquad (3)$$

and

$$c_{rot} = 1 + \frac{5}{16}\left(\frac{R}{h}\right)^3. \qquad (4)$$

Note: for Eq. (3), the translation direction is parallel to the plane. As the gap width $h - R$ approaches zero, lubrication theory [23] must be considered and the corrections are no longer valid. In such situations, the asperities on the sphere are in contact with the wall and it becomes likely that the no-slip boundary condition of the fluid no longer holds [23, 24]. At these surface proximities and low Reynolds numbers, shear-induced lift forces

are negligible [7] . Nevertheless, the contact between the particle and the surface may induce a rolling-while-slipping, or "skipping," behavior that is mediated by the surface roughness of either side, further complicating the situation.

There have been several theoretical descriptions of the rolling and slipping motion, down an inclined plane under the influence of gravity, of rough low-Reynolds number spheres in a viscous fluid [25, 26]. These treatments, however, do not deal with the motion of rotationally driven spheres, such as magnetically rotated microspheres. For such systems, nonlinear behavior arises when the external driving fields rotate at frequencies faster than can be supported by the rotational drag. For a magnetic microsphere aligning with an external rotating magnetic field, this nonlinear behavior emerges from an equation of motion involving a balance of torques given as

$$I\ddot{\theta} + \gamma_{rot}\dot{\theta} = mB\sin(\Omega t - \theta) \tag{5}$$

where $I$ is the moment of inertia of the particle, $m$ is the magnetic moment of the microsphere, $\Omega$ is the rotation frequency of the external magnetic field of strength $B$ and $\theta$ is again the phase angle of the microsphere with respect to the lab frame. Eq. (5) can be rewritten in dimensionless form by substituting $\Omega_c = mB/\gamma_{rot}$, $\tau = \Omega_c t$, and $\phi = \Omega t - \theta$. Furthermore, operation in the low Reynolds number regime [27] allows the inertial term to be neglected to obtain

$$\frac{d\phi}{d\tau} = \frac{\Omega}{\Omega_C} - \sin(\phi), \tag{6}$$

which is known as the nonuniform oscillator equation [4, 12-15]. Nonuniform oscillators described by Eq. (6) are characterized by the following solution for the average rotation rate:

$$\langle \dot{\theta} \rangle = \begin{cases} \Omega & \Omega \leq \Omega_c \\ \Omega - \sqrt{\Omega^2 - \Omega_c^2} & \Omega \geq \Omega_c \end{cases}. \quad (7)$$

The critical external frequency $\Omega_C$ is the frequency at which the oscillator abruptly transitions from being in-phase with the rotations of the driving field to being in a state in which the phase angle continually slips with respect to that of the external field.

At frequencies well below $\Omega_C$, an increase in the driving frequency would be accompanied by an increase in the microsphere rotation rate and thus a proportionate increase in the drag. At a surface, this effective drag torque is the rotational viscous drag of the solution given by Eq. (2) and Eq. (4), combined with frictional drag due to the surface. This additional friction may be caused by interactions between the particle and the surface such as electrostatic and van-der-Waals forces, as well as the roughness of either the microsphere or the surface. If the microsphere is rolling or skipping, the increased rotation rate results in a higher rolling velocity. At driving frequencies above $\Omega_C$, the average rotation rate decreases because the microsphere cannot keep up with the magnetic field. The microsphere continually slips with respect to the external field and, at a surface, even lower rolling velocities do result. When a magnetic microsphere is optically trapped at a surface, similar behavior is expected. At low external rotation frequencies, the proportionately increasing surface drag-induced torque displaces the microsphere further from the trap center. At even lower frequencies, the average rotation rate decreases along with the effective surface drag torque and the microsphere finds a balance closer to the trap center. For displacements smaller than the particle radius, we expect a Hookian (linear) dependence of the restoring force on the displacement. Thus, the functional dependence on frequency of the effective surface drag has the same qualitative form as position vs. frequency.

## Preliminary Results

To establish appropriate ranges for experimental parameters such as laser power and rotation rate, aminated microspheres on uncoated cover slides were first tested. Other parameters such as alignment angle of the rotation axis were also maintained throughout the remainder of the experiments. The preliminary results shown here illustrate the method of analysis, and procedures were representative of all experiments conducted.

### Free-rolling Microspheres

With the laser off, microspheres rolled along an almost-linear trajectory on an uncoated glass surface while rotated by an external magnetic field, either clockwise or counterclockwise. Figure 4 shows the linear displacement along the $y$-axis, of a $9.0 \pm 0.2$ $\mu m$ aminated magnetic microsphere, from its original position as a function of time, for several magnetic rotation frequencies. The displacement was measured along the $y$-axis, a direction approximately perpendicular to the axis of rotation, determined as follows. Along the axis of rotation, the microsphere typically remained less than 0.5 $\mu m$ from its original position at $t = 0$. Occasionally, at higher frequencies (~3 $Hz$), instead of rolling along the same direction as at lower frequencies, the trajectory would have a significant component along the rotation axis, sometimes deviating up to 15° from the low frequency trajectories in these preliminary results. The *y*-axis was thus determined as the direction that the microsphere rolled at the lowest measurable non-zero rotation rate. The rolling velocities, determined from the fitted slopes shown in Figure 4, increased with frequency until a threshold was reached. A summary of these slopes is provided in Figure 4(b). At higher rotational frequencies the rolling velocity decreased along the *y*-axis and was sometimes associated with slightly increased velocities along the *x*-axis.

For a 9 *μm* diameter particle at a rotation rate of 0.5 *Hz*, rolling without slipping requires a translational velocity of about 14 *μm/s*. Our particle rolled at 1 *μm/s*, thus even at rotation rates below the threshold for reduced rolling velocity, the microsphere was skipping. At or near the surface, even in the presence of SDS, the particles experienced some drag component that was frequency-dependent. This does not necessarily imply that the viscosity was non-Newtonian, just that the rolling resistance consisted of the standard Stokes drag [Eqs. (1)-(4)] as well as a frequency-dependent friction induced by surface interactions.

**Optically Trapped Microspheres**

Similar experiments were conducted for microspheres trapped by optical tweezers. Instead of rolling across the surface, the microspheres were spatially confined by the optical potential well. In such cases, the microsphere was displaced from the trap center to a position at which the trap restoring force counteracted the surface drag force that would have, otherwise, induced rolling.

Overlaid, averaged image stacks for the same $9.0 \pm 0.2$ $\mu m$ magnetic microsphere, manipulated at different rotational frequencies and laser powers, are shown in Figure 5. For higher laser powers, the trap stiffness was larger, causing the particle to be pulled closer to the center of the trap. This is visually indicated by the increased area of the region of overlap between rings. At a given laser power, as the rotation frequency was increased, the microsphere moved farther from the trap center. A threshold was reached, i.e. at 3.0 *Hz*, at which point increased slipping (reduced rolling friction) at the glass surface caused the particle to reside closer to the trap center. This can be visualized by the decreasing area of overlap with increasing rotational frequency, until the threshold at which point the area of overlap has slightly increased. This behavior was repeatable and reversible by adjusting the magnetic rotation frequency.

To quantify these results, the *x*- and *y*- coordinates of the microsphere image were tracked. As in the case of free-rolling, microspheres trapped and rotated at larger magnetic frequencies would experience forces along the rotation axis. Deviations from the axis perpendicular to the rotation axis were as high as 7° in preliminary results. The direction of the *y*-axis was again determined by the direction of displacement for the lowest measurable non-zero rotation rate. At zero-frequency, microspheres can settle into preferential orientations due to protrusions at the surface of the particle. Because of stiction, the particle position may not coincide with the center of the trap.

Histograms of the center coordinate of the trapped microsphere, adjusted to the appropriate coordinate system, were calculated using MATLAB (The Mathworks, Inc) analysis functions. These histograms are shown in Figure 6 for a laser power of 5 *mW*. The graphs in the top row (a, b) are for one rotation direction, the bottom row (c, d) for the opposite direction. The trap is located at the origin ($x = 0, y = 0$). The graphs in Figure 6 along with the images of Figure 5 verify that the rotational drag, due to the microsphere slipping at the surface, induced an overall positional shift in the y-direction, while the average displacement in the *x*-direction remained closer to the trap center.

Using MATLAB's nonlinear least-squares fitting routine *nlinfit*, each histogram was well-fitted to a Gaussian profile of the form:

$$n(x_i) = a_1 \exp(-\frac{(x_i - a_2)^2}{a_3^2}) \qquad (8)$$

where $a_i$ are the fit parameters and $x_i$ is either the *x*- or *y*-displacement. The Gaussian form indicates the microsphere experienced a normal (Boltzmann) distribution of positions even though the particle was displaced by the magnetically-induced rolling friction of the surface. Thus, in terms of the temperature of the solution *T*, this Gaussian distribution can also be written as:

$$n(x_i) = a_1 \exp(-\frac{\kappa(x_i - a_2)^2}{2k_B T}), \qquad (9)$$

where $\kappa$ is the effective radial spring constant of the system and $k_B$ is Boltzmann's constant. The width of this distribution tended to be larger than that resulting from damped Brownian motion of the particle trapped in the optical harmonic potential well without magnetic rotation. This suggests that interactions with the surface can be represented as a reduction in the effective spring constant of the optical trap ($\kappa = \kappa_{opt} - \kappa_{surface}$). The maximum displacement in all preliminary results was less than 4 $\mu m$ – a distance smaller than the particle radius. We expect the restoring force of the optical trap to obey Hookian dynamics, in other words, behave linearly with displacement. Furthermore, the dependence of the central position of the microsphere on rotational frequency proportionately reflects the dependence of the total effective drag torque on the microsphere due to the combined surface frictional torque and the torque of the surrounding fluid.

A plot of the center *y*-displacement from the trap for different magnetic rotational frequencies and laser powers is shown in Figure 7. The errorbar for each point, corresponding to the estimated standard deviation for the fitted position coefficient $a_2$ in Eq. (8), is smaller than the diameter of each marker with the maximum standard deviation over all measured displacements being 0.018 $\mu m$. For a given laser power, at low magnetic rotation frequencies, the microsphere resided close to the trap center. As the magnetic rotation rate was increased, the microsphere's average position shifted away from the trap center. At these farther distances, the increased rotational drag due to the presence of the surface balanced the increased restoring force of the optical trap. At frequencies near 2-2.5 *Hz* and above, the surface drag reached a threshold at which point the microsphere, while still slipping, was stably pulled closer to the trap. Video images revealed that the microsphere remained in focus at its equator to within 0.5 $\mu m$ (the minimum change in the *z*-focus that produced a noticeable change in observed focus). This suggests that contact with the surface persisted, at least intermittently, but the effective rotational frictional drag coefficient at the surface was frequency dependent, being reduced at higher frequencies.

These preliminary results provided the appropriate range of laser powers and rotational rates needed to observe the effect of interactions between the glass surface and a rotating magnetic microsphere. Such interactions are determined by the physical properties of both the substrate and the particle. We investigated both of these aspects by comparing the motion dynamics for the following modifications: (1) blank glass cover slides vs. cover slides coated with PEG and (2) rough carboxylic magnetic microspheres vs. smoother amine-functionalized magnetic microspheres. In addition, we increased the resolution of rotation frequencies in an effort to observe more closely the dynamics near the threshold for reduced friction.

## Results and Discussion

### Glass PEGylation vs. Non-PEGylation

Amine-functionalized magnetic microspheres were rolled along glass cover slides, both with and without PEG functionalization, at various rotation rates. Microspheres from the same batch were also rotated while held in an optical trap with an incident laser power of 3 $mW/cm^2$. The results of these two experiments are summarized in Figure 8(a) and (b), respectively, in which the magnitude of the measurements for clockwise and counter-clockwise rotation have been averaged. To account for microsphere size, velocities for free-rolling particles were normalized by the circumference at the equator and displacements for trapped particles were normalized by the radius of the particle.

Without the optical trap, for frequencies larger than 1 *Hz*, the rolling speed along the *y*-axis was larger for microspheres on blank cover slides than for those on PEGylated ones. For blank slides, the dependence on frequency indicates a sharp discontinuity near 2.5 *Hz*. For frequencies below this, the rolling speed increases proportionately with the magnetic rotation rate while for larger frequencies the rolling speed decreases with rotation rate. This behavior is similar to the dynamics of nonlinear oscillators, far from an interface, described by Eq. (7). Below the critical frequency $\Omega_C$ of such an oscillator, the particle is phase-locked, rotating synchronously with the external rotating magnetic field. The dashed line in Figure 8(a) shows the similarity in rotational response, of the amine-functionalized microsphere on an untreated glass cover slide, to that of a nonlinear oscillator. A reduction factor was introduced on the right hand side of Eq. (7) to account for the presence of the surface. The equation was then linear-least-squares-fit to the data, giving a critical frequency $\Omega_C = 2.48$ *Hz* and a reduction factor of 0.060. This reduction factor, which quantifies the continual slipping of the microsphere even below the apparent critical frequency, can be understood as the ratio of the normalized rolling speed to the external rotation rate in the low-frequency regime. A ratio equal to one would occur for rolling without slipping while much smaller ratios indicate significant skipping. A linear fit of the first three points of the corresponding data gives a slope of normalized rolling speed/rotation rate of 0.051 and an intercept of 0.007 *Hz* at $\Omega = 0$. We expect both the intercept to be non-zero and the slope of the data in the low frequency linear regime to be slightly less than our nonlinear oscillator reduction factor since the microsphere must transition from non-slip-rolling to skipping. Our results indicate that this difference in slope is negligibly small in our system, suggesting that the transition occurs at very low rotation frequencies.

The critical frequency of these amine-functionalized magnetic microspheres that are away from the surface could not easily be experimentally determined since the microspheres appeared to rotate in phase-locked fashion even at frequencies as high as ~10 $Hz$ (the mechanical limit of our magnetic rotation system). The critical frequency, for systems in water, can, however, be estimated, based on measurements performed on dimers of the same magnetic microspheres rotated in a 50% w/w glycerol/water mixture (data not shown). Using these results, which gave $\Omega_C$ = 1.3 $Hz$, along with the rotational drag coefficient for a dimer, $\gamma_{rot,dimer} = 29.92\pi\eta R^3$ [28], and the ratio of viscosities between the 50% mixture and pure water, $\eta_{50\%}/\eta_{water} = 5.21$ [29], we expected the critical frequency to be about 25 $Hz$. For our microspheres at the surface, this critical rate drops to 2.48 $Hz$, indicating the additional drag due to the presence of the surface. In addition, for standard nonlinear oscillators, Eq. (7) predicts a much steeper decrease in rotation rate just above the critical frequency. The more shallow decrease of our data suggests that the frequency-dependence of the interaction with the surface is more complicated than expected from a pure Stokes drag.

For PEG-coated slides, the rolling speed was always less than that for blank slides. This is expected since the interaction between the microspheres and blank cover slides was higher, as evidenced by a stronger non-specific binding with the blank cover slide surface. The PEG coating inhibits this binding effect and, as a result, the microspheres cannot gain as much traction with the surface. Consider the following simplified model of individual magnetic colloids interacting with the glass surface as the microsphere rotates: For low rotation frequencies, there is some tendency for individual magnetic colloids to interact with the glass surface through adhesion or binding. During some of these colloid-surface interaction events, the microsphere rolls forward. An increase in rotation rate corresponds to an increased number of interactions, resulting in a faster rolling speed. Beyond a certain frequency threshold, however, the duration of interaction between a colloid and the surface is decreased, thus reducing the likelihood that torque-generating events can occur. Asperities on the microsphere may experience more total contact with the surface, but less actual events that induce rolling. The particle slips more often resulting in slower rolling speeds at high rotation frequencies.

The rolling speed decreased with increasing rotation rate on PEGylated slides, although a slight discontinuity appears near 2 *Hz*. This discontinuity occurs at the same frequency as the sharp threshold corner in the results for the experiments conducted with microspheres trapped by optical tweezers, as shown in Figure 8(b). The frequency threshold for free-rolling amine-functionalized microspheres on blank slides also matches that for optically trapped ones. With the optical trap present, the same two-regime behavior occurs. At low frequencies, the displacement from the trap center increased with rotation rate, while at higher frequencies, beyond a certain threshold, the microsphere was pulled closer to the trap center. In this case, the position information corresponds to actual frictional forces that acted on the microsphere, in opposition to forces due to the optical trap. The evident disparity at low frequencies between free-rolling microspheres on PEGylated slides (decreasing speed) and optically trapped microspheres on PEGylated slides (increasing speed) can be explained as follows: When optical tweezers were applied to the microspheres, the laser power was first adjusted such that the axial scattering force on the particle was reduced enough to prevent microsphere levitation off the surface. Even at these reduced powers, the gradient force was strong enough to radially trap the particle. The axial gradient force could also pull the particle downwards, thus maintaining slipping-contact between the glass surface and the magnetic colloids on the microsphere, and inducing the microsphere to roll away from the trap. In the absence of the optical trap, no such load exists and the microsphere can shift upwards away from the surface, enough to cause increased slipping, thus causing the microsphere to roll away from its origin at a slower speed.

One should notice that, at low frequencies, the trend for optically trapped microspheres, as well as microspheres free-rolling on blank glass, indicates a linear increase with frequency. Extrapolation of this data to zero frequency again yields a positive rolling speed or displacement. Such results indicate that the microsphere does transition from non-slip rolling, which would yield a linear trend exhibiting a slope of unity, to rolling with slipping.

Differences in the dynamic motion of our microspheres due to glass PEGylation were also observed when analyzing the Gaussian distribution of positions of the optically trapped microspheres. The widths of the distributions, represented by the parameter $a_3$ given in Eq. (8), were generally larger on PEGylated slides than on uncoated slides. On PEGylated slides, the Gaussian widths for magnetically rotated microspheres were about 66% larger along the $x$-direction and 79% larger along the $y$-direction, compared to microspheres without magnetic rotation ($\Omega = 0$). In contrast, on uncoated glass cover slides, the Gaussian widths for magnetically rotated microspheres were about 29% larger and 47% larger for the $x$- and $y$- directions, respectively, compared to microspheres without magnetic rotation. Therefore, on PEGylated slides, the effective spring constant in Eq. (9) was reduced less than on uncoated slides, which makes intuitive sense.

Another noteworthy feature in the data represented in Figure 8 is that both the speeds of free-rolling microspheres and the distances from the trap center of optically trapped microspheres along the $x$-axis are not zero. Although small in comparison to corresponding values along the $y$-axis, these non-negligible measurements indicate that the trajectories were not purely perpendicular to the rotation axis, but also had components along the rotation axis. The magnitude of the angle of deviation from the $y$-axis, of these trajectories, averaged over both rotation directions, is shown in Figure 9. It has been shown that nanoscale objects with geometrically distributed facets rolling on commensurate surfaces can exhibit off-angle trajectories [30]. In our system, however, the microspheres appeared to have randomly distributed magnetic colloids (see Figure 1) and the glass surfaces were also expected to have a random distribution of potentially adhesive contact points. The observed increased deviation angles at higher rotation rates suggests that the magnetic moment of the microsphere, no longer remained in a plane perpendicular to the rotation axis, instead of aligning with the magnetic field by rotating end-over-end. This escape into the third dimension has been described in other magnetically driven systems [10, 31, 32].

**Particle Roughness**

Two distinct types of magnetic microspheres were rotated on glass cover slides, again with and without glass PEGylation, at varying external rotation rates. As shown in the SEM images in Figure 1, carboxylated microspheres appeared to have more magnetic material and an increased surface roughness, in comparison to amine-functionalized ones. In addition, the carboxylated microspheres tended to irreversibly bind onto the uncoated cover slides, if the rotation rate was below 1.5 *Hz*. The effect of microsphere roughness on the hydrodynamic motion perpendicular to a plane (sedimentation) has been studied in great detail [33, 34]. We sought to compare the effects of microsphere roughness for the motion parallel to a surface. Results for the carboxylated microspheres, free-rolling and confined by an optical trap, are shown in Figure 10(a) and (b), respectively. These results will be compared to the measurements on amine-functionalized microspheres shown in Figure 8(a) and (b), where appropriate.

In all the cases of free-rolling microspheres studied here, the normalized rolling velocities on blank cover slides were generally faster than those of the same type microsphere on PEGylated cover slides, at the same rotation rate. The increased traction on blank slides enabled the microspheres to experience increased torques and thus roll faster along the surface. For the rougher carboxylated microspheres, the ratio of normalized rolling speed to rotation rate, given by the slope of the linear fit of the data in Figure 10(a), was 0.050 for the blank cover slide. This value is similar to that calculated for the aminated microspheres on blank cover slides; thus, in the low frequency regime, the slipping friction is approximately the same as for untreated slides. For the PEG-coated cover slides, however, the rougher carboxylic particles introduced slightly more friction, giving a positive slope of 0.045, as opposed to the negative slope of the smoother amine-functionalized microspheres, which rolled slower with increasing rotation rate.

This last difference mentioned suggests another behavioral distinction in the rolling motion of the rougher carboxylic microspheres vs. smoother amine-functionalized ones. The sharp threshold between increasing and decreasing rolling speeds, typically signaling the transition from phase-locked rotation to phase-slipping rotation, occurred at significantly higher rotation rates for the rougher carboxylic microspheres, compared with the smoother amine-functionalized ones. With nonlinear oscillators in the bulk fluid, increased drag does shift the threshold frequency $\Omega_C$ to lower frequencies. Alternatively, at a surface, an increase in the surface drag shifts the threshold for the rolling speed to higher frequencies, since this drag provides the traction necessary to induce rolling. We see that, for PEGylated cover slides, the carboxylated microspheres reached the threshold near 4 *Hz,* while, on blank cover slides, the rolling speed continued to increase over all frequencies measured, despite the increase in angular deviation suggested by the increased speed along the *x*-direction.

For optically trapped, carboxylated magnetic microspheres, the results were not as conclusive. Figure 10(b) indicates the average distance from the trap center for varying rotation rates. The data appears to fluctuate significantly, revealing no clear sign of a threshold or suggestive trend. Several factors are attributed to this contrast in results. First, since the carboxylated microspheres had characteristically more magnetic material than the aminated ones, scattering forces were increased. Thus, a lower laser intensity, of only 1 $mW/cm^2$, was necessary to prevent the scattering force from lifting the microspheres off the glass surface. This weaker laser power also reduced the gradient force that was necessary for trapping the microspheres. It is possible that the carboxylated microspheres experienced a reduced load since the gradient force pulling downward was weaker. Thus, the microspheres were more likely to fluctuate in the $z$-direction, causing less contact with the surface and, overall, less effective surface drag. Second, large asperities on the surface of the carboxylated microspheres caused the microsphere to shift position laterally while rotating in the optical trap, thus causing a large error in several measurements, indicated in Figure 10(b). This may have been a combination of skewed orientation on the glass surface as well as small optically induced rotation, caused by the anisotropy of absorptive aggregates within the shell of magnetic colloids. It is conceivable that such fluctuations could cause errors in the normalized distance from the trap center of up to 0.05. Last, the carboxylated microspheres had a tendency to bind to blank cover slides much more readily than the aminated ones, especially at very low frequencies. Consequently, measurements below 1.75 $Hz$ could not be acquired before the microspheres irreversibly bound to the surface.

Angular deviations for both free-rolling and optically trapped carboxylic microspheres are shown in Figure 11. Results for each rotation direction were averaged. The deviation is not as strong as for the amine-functionalized magnetic microspheres. This suggests that the interactions between the surface and the rougher, more adhesive, carboxylic microspheres were suppressing tendencies for the magnetic moment to rotate out of plane. In addition, the data appears noisier, which we suspect is another consequence of the numerous asperities found on the microsphere surface.

For both amine-functionalized and carboxylated magnetic microspheres that are rotating while optically trapped, we've neglected effects due to changes in temperature, resulting from the absorption of laser light. Absorption by the surrounding aqueous solution is negligible at such low laser intensities [35], since temperatures are expected to increase less than $0.05°\ K$ and, furthermore, the index of refraction of water is only weakly dependent on temperature [36]. Absorption by the magnetic colloids, however, followed by heat transfer to the surrounding fluid, could reduce the fluid viscosity. This effect may contribute to the scatter in the data for optically trapped carboxylic magnetic microspheres.

## Conclusions and Future Work

We have developed an experimental technique for studying the motional behavior of rotationally driven magnetic microspheres that are rolling while slipping on a planar glass surface. Our results justify the notion that both the surface roughness effects and those due to hydrodynamic interactions between planar surfaces and different individual microspheres can be distinguished and compared. This capability of measuring the adhesion and binding properties of rolling objects is of great interest in applications such as the fabrication of cell sorting and cell separation microfluidic [9, 37] and optofluidic [38] devices. In particular, microspheres and surfaces with increased drag characteristics exhibit higher microsphere rolling speeds and larger deviations from the center of an optical trap at low rotational frequencies. At higher frequencies, increased surface-slipping can occur and a behavior characteristic of nonlinearly oscillating, driven magnetic microspheres can be observed, namely escape of the magnetic moment into the third dimension, thus causing off-angle surface rolling.

A more detailed theoretical study is required to predict the hydrodynamic motion of rotationally driven microspheres on rough surfaces. Studies have previously been conducted on the rotational and translational behavior of rough non-colloidal spheres that are pulled by gravity down inclined planes [25, 26]. It is unclear, however, whether similar theoretical treatments involving the balance of drag forces and torques can be applied to these rotationally-driven rough microspheres rolling and slipping on flat planes.

If one could measure the magnetic moment of a single particle, this would help distinguish whether variations in magnetic content from microsphere to microsphere versus the effective drag are contributing to shifts in the threshold frequency for different surfaces. Such measurements are straightforward using optically anisotropic probes such as half-coated microspheres or dimers of particles [13]. For the microspheres used in the experiments described here, incorporation of optical anisotropy is ideal as long as the particle surface roughness remains unaffected. Also, precise measurement of the height of the microsphere with respect to the glass surface in real time, for example by using total internal reflection microscopy techniques [33, 39], could elucidate the phenomenon of rolling while slipping, by verifying when microsphere-surface contact occurs.

**Acknowledgments**

The authors thank Vladimir Stoica and Prof. Roy Clarke for use of their Gauss Probe, Kate Dooley and Prof. Michael Morris for assistance in PEGylation characterization, and Gwangseong Kim for assistance with SEM characterization. SEM images were acquired at the Electron Microbeam Analysis Laboratory (EMAL) at the University of Michigan. This work was funded by NSF grant DMR 0455330 (RK).

[1]     B. Bhushan, J. Israelachvili, and U. Landman, "Nanotribology: Friction, wear and lubrication at the atomic scale," *Nature* **374**(6523), 607-616 (1995).

[2]     S. Kim, D. Asay, and M. Dugger, "Nanotribology and MEMS," *Nano Today* **2**(5), 22-29 (2007).

[3]     I. Gebeshuber, "Biotribology inspires new technologies," *Nano Today* **2**(5), 30-37 (2007).

[4]     K.D. Bonin, B. Kourmanov, and T.G. Walker, "Light torque nanocontrol, nanomotors and nanorockers," *Opt. Express* **10**(19), 984-989 (2002).

[5]     Z. Cheng, T. Mason, and P.M. Chaikin, "Periodic oscillation of a colloidal disk near a wall in an optical trap," *Phys. Rev. E* **68**(5), 051404 (2003).

[6]     Z. Burton and B. Bhushan, "Hydrophobicity, adhesion, and friction properties of nanopatterned polymers and scale dependence for micro- and nanoelectromechanical systems," *Nano Lett.* **5**(8), 1607-1613 (2005).

[7]     R. Duffadar and J. Davis, "Interaction of micrometer-scale particles with nanotextured surfaces in shear flow," *J. Colloid Interf. Sci.* **308**(1), 20-29 (2007).

[8] D.A. Hammer and S.M. Apte, "Simulation of cell rolling and adhesion on surfaces in shear flow: General results and analysis of selectin-mediated neutrophil adhesion," *Biophys. J.* **63**(1), 35-57 (1992).

[9] S. Hong, D. Lee, H. Zhang, J.Q. Zhang, J.N. Resvick, A. Khademhosseini, M.R. King, R. Langer, and J.M. Karp, "Covalent immobilization of P-selectin enhances cell rolling," *Langmuir* **23**(24), 12261-12268 (2007).

[10] K. Erglis, Q. Wen, V. Ose, A. Zeltins, A. Sharipo, P.A. Janmey, and A. Cēbers, "Dynamics of magnetotactic bacteria in a rotating magnetic field," *Biophys. J.* **93**(4), 1402-1412 (2007).

[11] A. Cēbers and M. Ozols, "Dynamics of an active magnetic particle in a rotating magnetic field," *Phys. Rev. E* **73**(2) (2006).

[12] S.H. Strogatz, *Nonlinear Dynamics and Chaos*. 1994: Addison-Wesley Reading, MA.

[13] B.H. McNaughton, R.R. Agayan, J.X. Wang, and R. Kopelman, "Physiochemical microparticle sensors based on nonlinear magnetic oscillations," *Sens. Actuators B* **121**(1), 330-340 (2007).

[14] G. Helgesen, P. Pieranski, and A.T. Skjeltorp, "Nonlinear phenomena in systems of magnetic holes," *Phys. Rev. Lett* **64**, 1425-1428 (1990).

[15] W.A. Shelton, K.D. Bonin, and T.G. Walker, "Nonlinear motion of optically torqued nanorods," *Phys. Rev. E* **71**(3), 036204 (2005).

[16] B. Lin, J. Yu, and S. Rice, "Direct measurements of constrained Brownian motion of an isolated sphere between two walls," *Phys. Rev. E* **62**(3), 3909 (2000).

[17] R. Schlapak, P. Pammer, D. Armitage, R. Zhu, P. Hinterdorfer, M. Vaupel, T. Fruhwirth, and S. Howorka, "Glass surfaces grafted with high-density poly(ethylene glycol) as substrates for DNA oligonucleotide microarrays," *Langmuir* **22**(1), 277-285 (2006).

[18] A. Ashkin, "Forces of a single-beam gradient laser trap on a dielectric sphere in the ray optics regime," *Biophys. J.* **61**(2), 569-582 (1992).

[19] A. Ashkin, J.M. Dziedzic, J.E. Bjorkholm, and S. Chu, "Observation of a single-beam gradient force optical trap for dielectric particles," *Opt. Lett.* **11**(5), 288-290 (1986).

[20] A. Ashkin, "Acceleration and trapping of particles by radiation pressure," *Phys. Rev. Lett.* **24**(4), 156-159 (1970).

[21]  J.P. Gordon, "Radiation forces and momenta in dielectric media," *Phys. Rev. A* **8**(1), 14-21 (1973).

[22]  B.C. Carter, G.T. Shubeita, and S.P. Gross, "Tracking single particles: A user-friendly quantitative evaluation," *Physical Biology* **2**(1), 60-72 (2005).

[23]  A.J. Goldman, R.G. Cox, and H. Brenner, "Slow viscous motion of a sphere parallel to a plane wall--I Motion through a quiescent fluid," *Chem. Eng. Sci.* **22**(4), 637-651 (1967).

[24]  A. Davis, M. Kezirian, and H. Brenner, "On the Stokes-Einstein model of surface diffusion along solid surfaces: Slip boundary conditions," *J. Colloid Interf. Sci.* **165**(1), 129-140 (1994).

[25]  K.P. Galvin, Y. Zhao, and R.H. Davis, "Time-averaged hydrodynamic roughness of a noncolloidal sphere in low Reynolds number motion down an inclined plane," *Phys. Fluids* **13**(11), 3108-3119 (2001).

[26]  J.R. Smart, S. Beimfohr, and D.T. Leighton Jr., "Measurement of the translational and rotational velocities of a noncolloidal sphere rolling down a smooth inclined plane at low Reynolds number," *Phys. Fluids A* **5**(1), 13-24 (1993).

[27]  E.M. Purcell, "Life at low Reynolds number," *Am. J. Phys.* **45**(1), 3-11 (1977).

[28]  T.G.M. van de Ven, *Colloidal Hydrodynamics*. 1989, New York: Academic Press Limited.

[29]  P.N. Shankar and M. Kumar, "Experimental determination of the kinematic viscosity of glycerol-water mixtures," *Proc. R. Soc. London, Ser. A* **444**, 573-581 (1994).

[30]  M.R. Falvo, J. Steele, R.M. Taylor, and R. Superfine, "Gearlike rolling motion mediated by commensurate contact: Carbon nanotubes on HOPG," *Phys. Rev. B* **62**(16), R10665 (2000).

[31]  C. Caroli and P. Pincus, "Response of an isolated magnetic grain suspended in a liquid to a rotating field," *Z. Phys. B Con. Mat.* **9**(4), 311-319 (1969).

[32]  R.R. Agayan and R. Kopelman, "Nonlinear 3-dimensional motion of a rigid dimer of magnetic microspheres in a rotating magnetic field," *Phys. Rev. E*, [To be Submitted] (2008).

[33]  J. Walz and L. Suresh, "Study of the sedimentation of a single particle toward a flat plate," *J. Chem. Phys.* **103**(24), 10714-10725 (1995).

[34]  J. Smart and D. Leighton, "Measurement of the hydrodynamic surface roughness of noncolloidal spheres," *Phys. Fluids A* **1**(1), 52-60 (1989).

[35]  E. Peterman, F. Gittes, and C. Schmidt, "Laser-induced heating in optical traps," *Biophys. J.* **84**(2), 1308-1316 (2003).

[36]  A.B. Djurisic and B.V. Stanic, "Modeling the temperature dependence of the index of refraction of liquid water in the visible and the near-ultraviolet ranges by a genetic algorithm," *Appl. Optics* **38**, 11-17 (1999).

[37]  H. Andersson and A. van den Berg, "Microfluidic devices for cellomics: A review," *Sens. Actuators B* **92**(3), 315-325 (2003).

[38]  A.H.J. Yang and D. Erickson, "Stability analysis of optofluidic transport on solid-core waveguiding structures," *Nanotechnology* **19**(4), 45704-45714 (2008).

[39]  M.A. Brown and E.J. Staples, "Measurement of absolute particle-surface separation using total internal reflection microscopy and radiation pressure forces," *Langmuir* **6**(7), 1260-1265 (1990).

**Figure Captions**

**Figure 1. Scanning electron microscope images of (a, c) amine-functionalized magnetic microspheres and (b, d) carboxylated magnetic microspheres. Images (a) and (b) show that the distribution of particle size was about 7-11 $\mu$m for both microsphere types. Amine-functionalized microspheres appeared to have less magnetic material and a decreased surface roughness compared to the carboxylated ones.**

**Figure 2.** Static sessile drop technique for distinguishing (a) uncoated glass from (b) PEGylated glass cover slides. The increased contact angle on PEGylated slides (59 ± 3°) compared to untreated slides (31 ± 4) was due to reduced wetting of the glass surface.

**Figure 3.** Aminated magnetic microspheres viewed under reflection mode microscopy. (a) Focused below the equator of the microsphere, magnetic colloids can be seen on the surface around a ring of focus. (b) Microsphere rotated by an external magnetic field rolling away from the center of an optical trap indicated by the blue circle. (c) Overlay of averaged images of the same optically trapped microsphere rolling due to clockwise and counter-clockwise magnetic rotation.

**Figure 4.** (a) Linear displacement perpendicular to the axis of rotation of a 9.0 ± 0.2 $\mu m$ aminated magnetic microsphere from its original position as a function of time for several magnetic rotation frequencies, both clockwise (+) and counterclockwise (-). (b) Rolling velocity magnitude increases with rotation frequency until a threshold is reached near 2 *Hz*. Above this threshold, the rolling velocity magnitude decreases.

**Figure 5.** Overlaid, averaged image stacks for an optically trapped 9.0 ± 0.2 $\mu m$ aminated magnetic microsphere manipulated at different rotational frequencies, both clockwise and counter-clockwise, and at varying laser powers. Displacement from the trap center was symmetric between both rotation directions. The **amplitude of the displacement can be visualized by comparing the area of overlap between rings. An increase in laser power pulls the microsphere closer to the trap center. An increase in rotational frequency displaces the trap further until a frequency threshold is reached, after which the microsphere resides closer to the trap center.**

**Figure 6. Histograms of the center coordinates of an aminated magnetic microsphere optically trapped at a surface by a 5 *mW/cm$^2$* laser beam and rotated by an external magnet. Graphs (a) and (b) are for one rotation direction while (c) and (d) are for the opposite direction. Rotational drag due to the microsphere rolling while slipping at the surface induced an overall positional shift away from the trap center at (0,0). The displacement along the *y*-axis was significantly larger than along the *x*-axis.**

**Figure 7.** Displacement from trap center for a rolling-while-slipping aminated magnetic microsphere optically trapped at a surface by varying laser powers and magnetically rotated at varying frequencies. Dotted lines indicate *x*-displacement while solid lines indicate *y*-displacement. Positive *y*-displacements occurred for clockwise rotation of the external magnet while negative *y*-displacements occurred for counter-clockwise rotations. In all cases, the microsphere displacement magnitude increases with rotation frequency until a threshold is reached. Above the threshold, increased slipping causes the microsphere to be pulled closer to the trap center.

**Figure 8.** (a) Normalized speed of amine-functionalized magnetic microspheres rolled along glass cover slides, both with and without a thin coating of PEG, at various magnetic rotation rates. Data points show averages over both rotation directions. Normalization is performed by dividing the rolling velocity by the circumference of the microsphere; therefore, normalized speeds less than one indicate rolling with slipping (skipping). The dashed line indicates the average rotation rate in *Hz* of a standard nonlinear oscillator multiplied by a reduction factor to account for friction due to the surface. (b) Normalized distance from the trap center of the same magnetic microspheres optically trapped by a laser beam with 3 *mW/cm²* of incident laser power on both uncoated and PEGylated glass cover slides. Normalization is performed by dividing the trapped distance by the particle radius. Data points show averages of each rotation direction.

**Figure 9.** Deviation angle for amine-functionalized magnetic microspheres (a) free-rolling and (b) optically trapped on either PEGylated (filled circles) or uncoated (open circles) glass cover slides.  Data points represent the average of the magnitudes over both rotation directions.  The increase in angle with rotation frequency suggests that the magnetic moment of the microsphere escapes into the third dimension along the rotation axis of the external magnetic field.

Figure 10. (a) Normalized speed of carboxylated magnetic microspheres rolled along glass cover slides, both with and without a thin coating of PEG, at various rotation rates. Data points are averages over each rotation direction. At low frequencies, for blank cover slides, the ratio of normalized rolling speed to rotation rate for carboxylated microspheres is similar to that for aminated ones. On PEGylated slides, the ratio increases for carboxylated microspheres in contrast to the decreasing ratio for amine-functionalized microspheres shown in Figure 8(a). The frequency threshold separating phase-locked rotation and phase-slipping rotation occurs at higher rotation rates for rougher carboxylic microspheres, compared with the smoother amine-functionalized ones. (b) Normalized distance from the trap center of the same magnetic microspheres optically trapped by a laser beam with 1 *mW/cm$^2$* of incident laser power on both uncoated and PEGylated glass cover slides. Significant fluctuations in the data are attributed to variations in the asperities of the magnetic colloidal shell of the rough carboxylic microspheres. A weaker trapping force was applied which induced less microsphere-surface contact, thus allowing further fluctuations in the data.

Figure 11. Deviation angle for carboxylated magnetic microspheres (a) free-rolling and (b) optically trapped on PEGylated and uncoated glass cover slides. The increase in angle with rotation frequency is not as strong as for smoother amine-functionalized microspheres. This suggests escape of the magnetic moment into the third dimension is partially suppressed by surface-microsphere interactions.

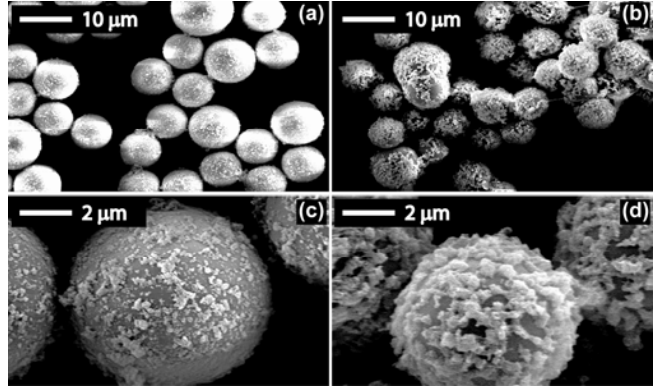

Figure 1(a)-(d).

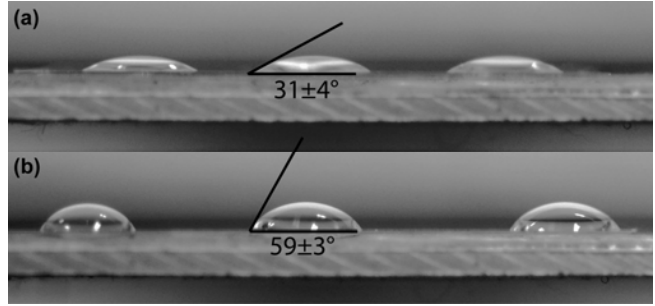

Figure 2(a) and (b).

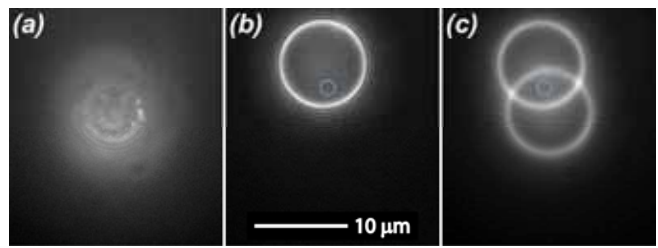

Figure 3(a)-(c).

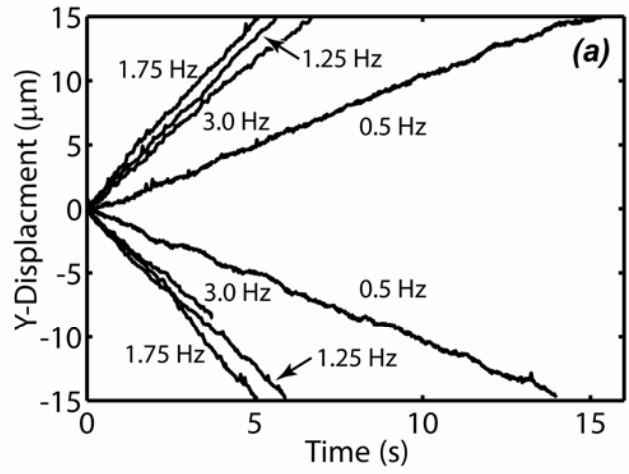
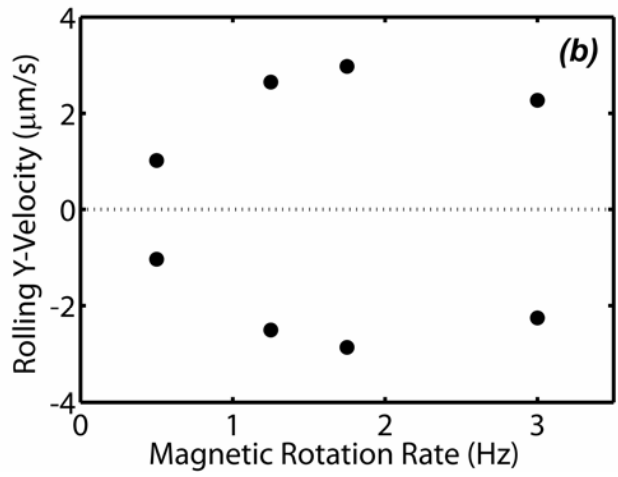

Figure 4(a) and (b).

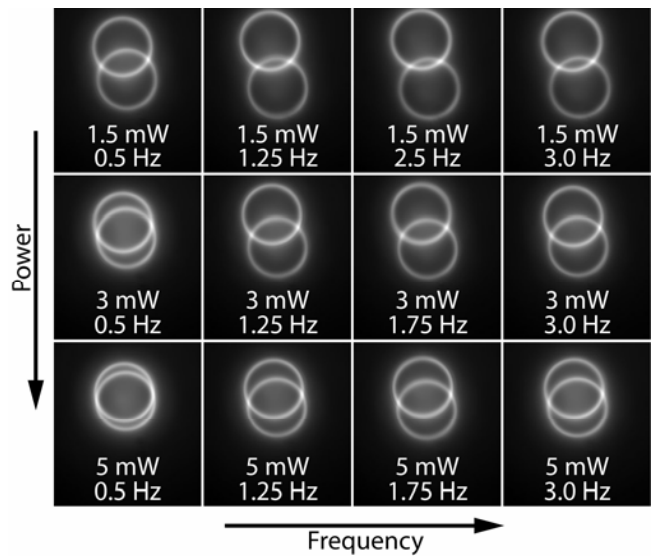

Figure 5.

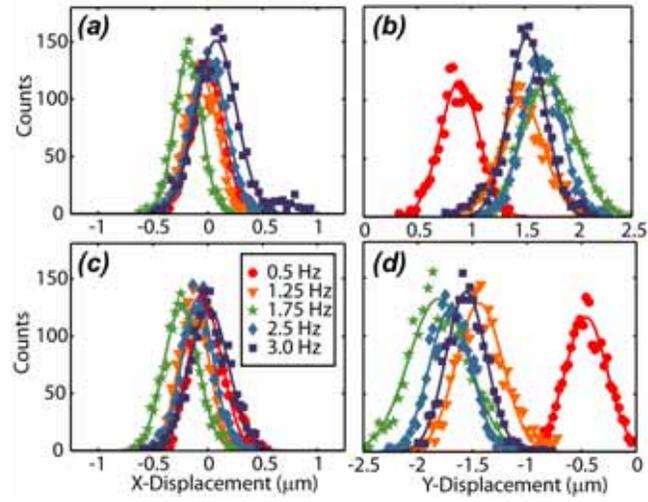

Figure 6(a)-(d).

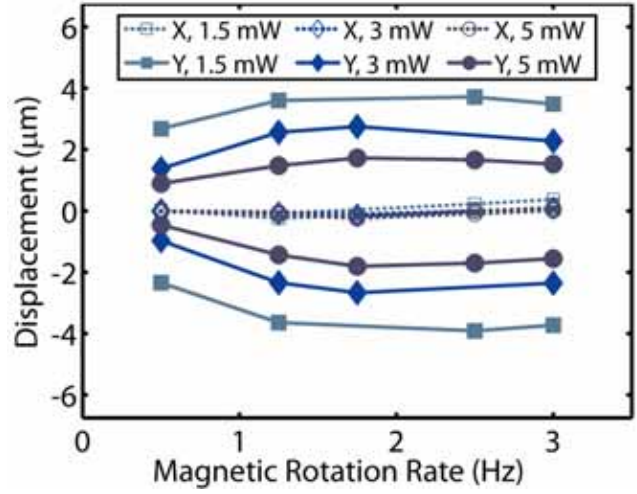

Figure 7.

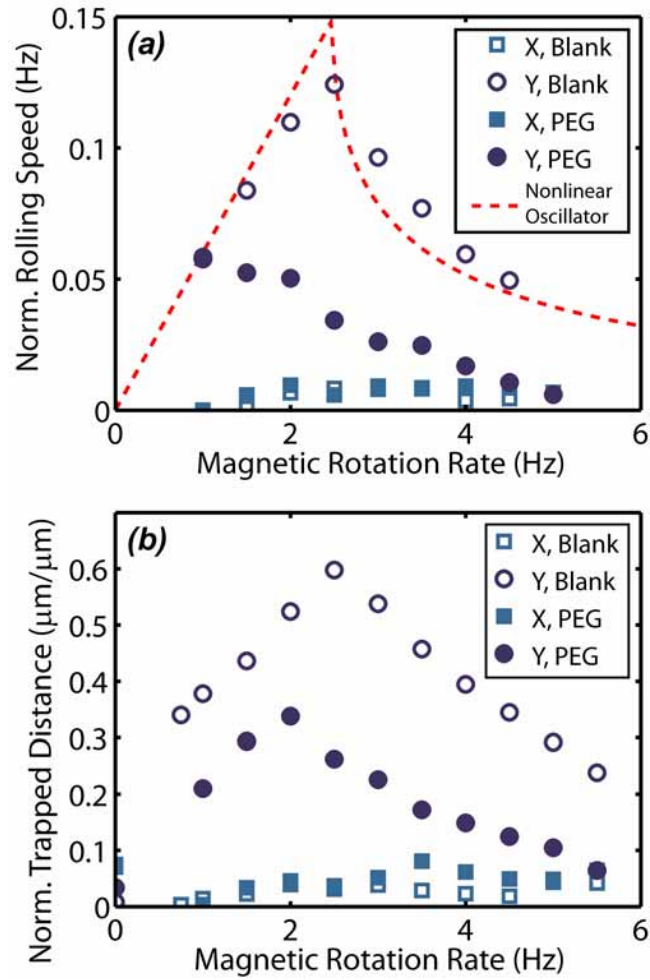

Figure 8(a) and (b).

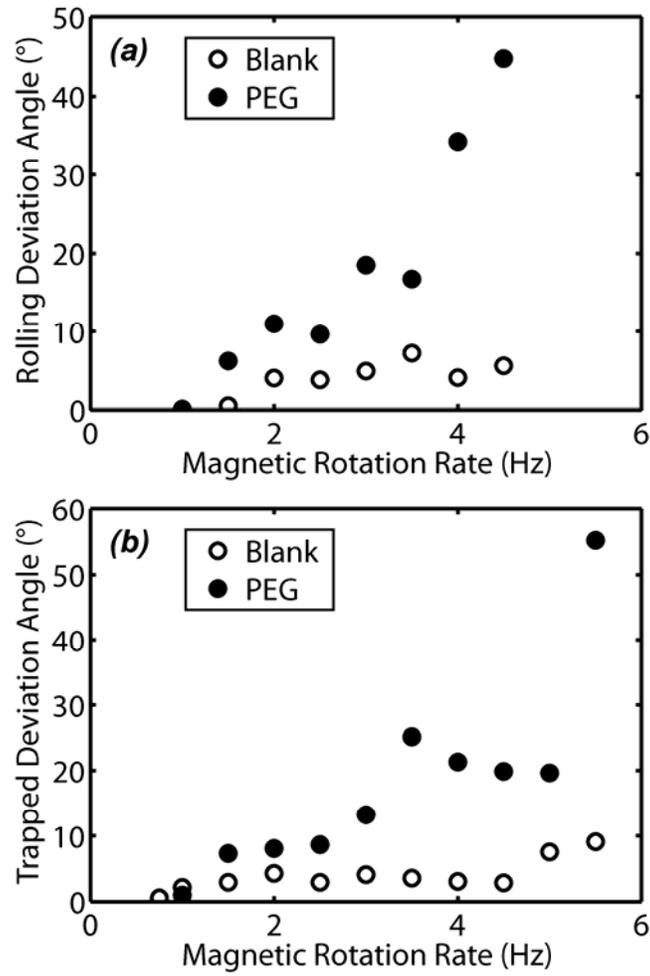

Figure 9(a) and (b).

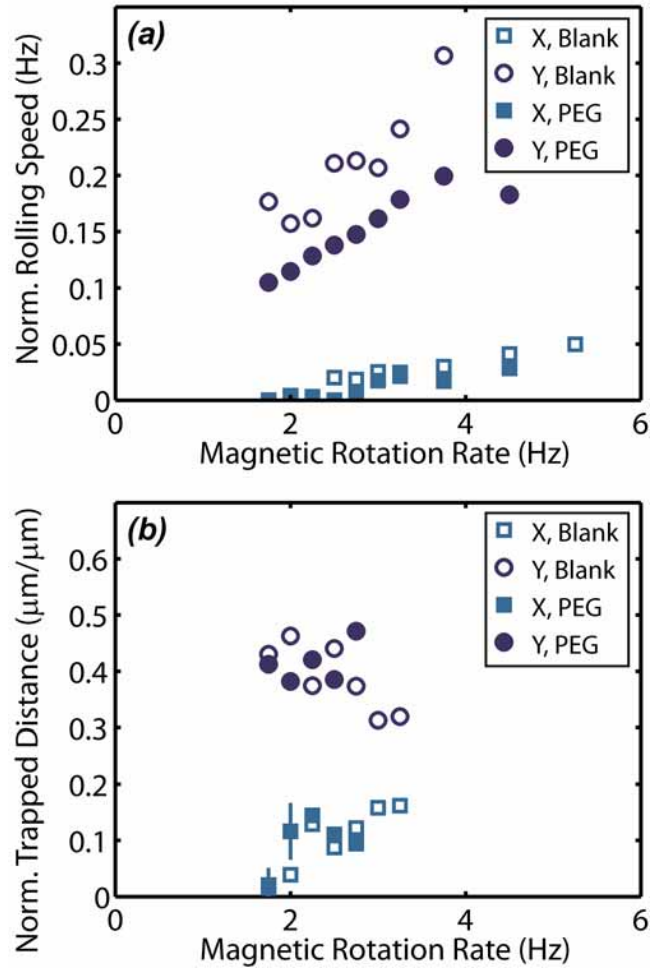

Figure 10(a) and (b).

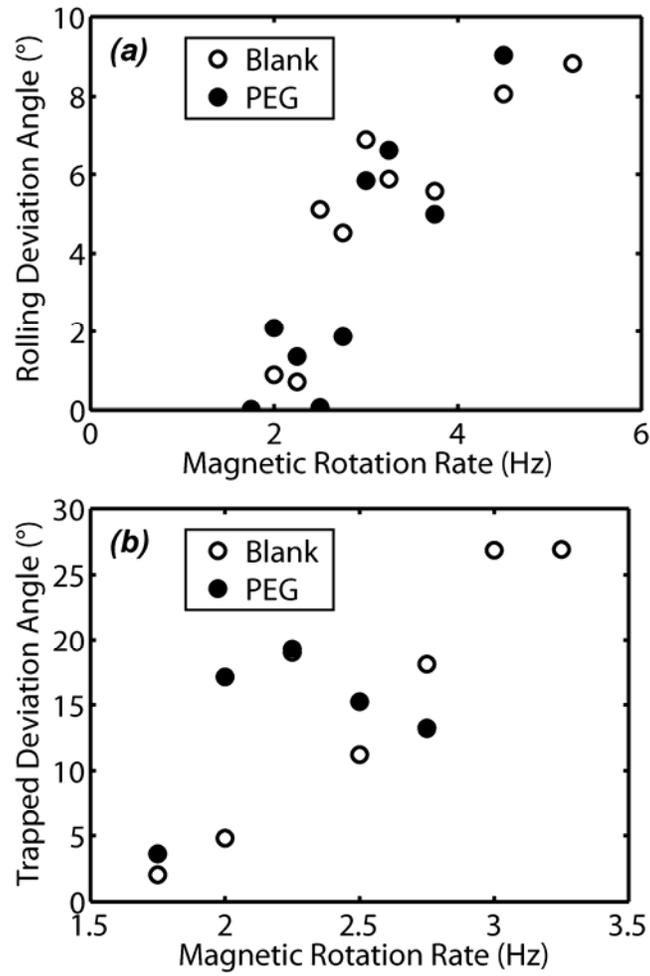

Figure 11(a) and (b).